# Synchronicity and pure bending of piezoelectric bimorphs: a new approach to kinetic energy harvesting

Michele Pozzi

School of Engineering, Newcastle University, Newcastle upon Tyne, NE1 7RU, UK

## Abstract

Kinetic energy harvesting with piezoelectric bimorphs has attracted considerable research interest in recent years. Many works have been dedicated to the modelling and optimisation of the cantilevered geometry to increase power density, bandwidth, etc. The increased efficiency coming from the use of trapezoidal beams has been recognised, but little has been done to produce the same uniform strain within the most commonly available rectangular beams. This work proposes a new approach via a smart compliant structure which permits to deform a set of bimorphs in pure bending. Furthermore, since the deflections are synchronous, the power signals produced are in phase and power conditioning is simplified and made more efficient. The kinematic requirements for uniform strain are discussed, the novel structure is proposed and modelled with finite elements, a prototype is presented and characterised to support the modelling. The proposed structure induces almost perfectly uniform strain in the piezoelectric beams for all useful rotation angles, demonstrating that, compared to a traditional cantilever, twice as many charges can be produced when the same maximum strain is applied to the material. Synchronicity is also experimentally verified for the prototype, as power signals resulting from impact excitation are observed to be in phase. The principle of synchronous pure bending via helper structures can be applied in general to increase the performance of piezoelectric energy harvesters.

## Introduction

The aim of this paper is to present a smart compliant structure for piezoelectric energy harvesting which ensures the synchronicity of power signals produced by an array of bimorphs, while also achieving pure bending to maximise the utilisation of the material.

Manufacturers of microelectronic components have been increasing their effort to reduce power consumption of devices. Starting a few decades ago, power requirements of microcontrollers have decreased to the point that it has become meaningful to seek to power small intelligent system by using stray energy present in the environment where they operate. Energy harvesting (or *scavenging* as it was more commonly called in the early days) has therefore developed into an independent area of research with the aim of enabling the deployment of energy autonomous systems. Due to the variety of stray energy present in different environments, a wide range of energy harvesting techniques and materials are being researched: thermoelectric generators [1], electrets [2], electromagnetic induction [3] and so forth.

Leveraging on the presence of movement in many environments, considerable effort is being devoted to kinetic energy harvesters, particularly using electromechanical transducers like piezoelectric devices.

There are many examples in the literature of works aimed at improving different aspects of kinetic energy harvesters. A range of low profile and small footprint devices are reviewed in

[4]. Recently, the response bandwidth of vibrational harvesters has attracted considerable interest: since environmental vibrations may fall over a wide range of frequencies, harvesting effectiveness can be increased with wideband devices. An array of beams, each with its own natural frequency and together covering the desired bandwidth, has been used in several studies [5,6]. Frequency tuning has been demonstrated by changing the stiffness of the cantilevered beam by exploiting magnetic interaction [7]. A wider operating band has also been achieved by introducing non-linear response and chaotic behaviour [8,9].

There are fewer works dedicated to improving the effective use of the piezoelectric material. The advantages of trapezoidal beams, and the uniform strain that can be achieved within them, have long been recognised for actuators [10] and for energy generation [11]. Often researchers have extended the vibrating beam by adding inert materials, extensions and extended seismic masses [12,13], with the intent of increasing the bending moment applied at the tip of the beam. The fabrication of more complex beams with varying thickness was also proposed to increase the strain uniformity and thence energy harvesting performance [14]. Air-spaced piezoelectric cantilevers have also been offered as an alternative to conventional bimorphs with the advantage of producing, if well designed, higher voltages and more uniform strain [15]. More recently, the principle of four-point bending has been applied to vibrational energy harvesting [16]. The common goal of the works just described is a greater uniformity of strain within the piezoelectric material.

The state of pure bending deformation is of interest in several other research areas and efforts are recorded to devise mechanisms that permit testing of materials in pure bending. In many cases, the mechanics is complex. In textile testing, the Kawabata Evaluation System for Fabrics (KES-F) is used. Relatively simple mechanisms have been devised to test materials of interest to cryogenics, although they only approximate the desired deformation [17,18]. More complex systems for biomechanical testing of spine segments have been proposed [19,20]. The testing systems for larger samples of generic shape are even more complex [21]. These few examples have been reviewed to highlight the difficulty of producing pure bending; when mechanisms are designed for testing of materials, high complexity is acceptable. For an energy harvesting application, the structure must be much simpler to minimise costs, dimensions, weight and energy dissipation. The present paper offers a novel approach to address this issue.

The majority of energy harvesters presented to date focuses on a single piezoelectric bimorph. However, in many applications there is an opportunity to place more transducers, which could work in concert to maximise the energy extracted from the environment. If they are active at the same time, the issue of synchronicity arises. Since the out-of-phase vibration of several transducers would yield a partial cancellation of the signal, with electrical energy produced by one re-injected into another, system efficiency is compromised. To avoid signal cancellation, in the past individual rectification has been added to each bimorph before further power management or storage [5,22].

The issue of synchronicity of power signals has come to the authors' attention when researching the Pizzicato series of energy harvesters. These are rotational harvesters featuring several piezoelectric bimorphs mechanically [23,24] or magnetically plucked [22]. Since the plucking events are uncorrelated, the signals generated by the bimorphs are in general out of phase. Individual biomorph's rectification was therefore necessary before further power management. This has several disadvantages, including cost and complexity of the electronics and losses in the numerous rectifying bridges. A better power management, based for example on synchronized switch harvesting on inductor (SSHI) or single supply pre-biasing (SSPB) [25], would be even more costly and it is impractical to have identical copies of such circuit for each of the many bimorphs an application might rely on. It is therefore necessary to devise a way to synchronize the outputs of the many piezoelectric devices in a rotational harvester so that they work in unison and provide outputs without phase shifts between them. The proposed solution relies on a ring onto which the tips of the bimorphs are attached, so that when this rotates with respect to a central hub (holding the roots of the beams), the vibrations of the bimorphs are synchronised. Whereas in the past [23] the issue of uniform strain was addressed by using trapezoidal bimorphs, the structure proposed here has the capability of bending rectangular beams into arcs, achieving near uniform longitudinal strain, together with synchronicity.

## Geometry for pure bending

Since excessive tensile stress levels reduce the operating life of ceramic piezoelectric transducers by introducing cracks and, eventually, causing failure of the material, it is worthwhile to consider the stress/strain distribution within a piezoelectric beam used for EH. A geometry where there is a stress concentration in one small area implies that to protect it from premature failure, other regions will be subject to lower than optimal stress levels. This reduces performance as the electrical output will be determined by an averaged value of strain. Considering bimorphs or monomorphs, which work in *31* mode, the ideal situation is to have a uniform stress/strain field along the beam (direction *1*); however, the traditional cantilever configuration has zero strain at the tip. Within a bending beam, the maximum strain will be observed at the external surfaces of the device, decreasing as we move inwards towards the neutral surface. The tensile strain ε at the convex external surface of a beam with rectangular cross section can be estimated as the product of half-thickness θ and curvature:

$$\epsilon(\xi) = \theta \frac{\frac{d^2 w}{d\xi^2}}{\left[1 + \left(\frac{dw}{d\xi}\right)^2\right]^{3/2}} \qquad 1$$

where $\xi$ is the longitudinal coordinate and $w(\xi)$ is the deflection of the beam from the relaxed horizontal position; all coordinates are adimensionalised. Note that for small deflections the omission of the denominator leads to an acceptable approximation.

Starting from standard equations found in textbooks for the shape of: a vibrating cantilever, a statically deflected cantilever and a beam bent into an arc, respectively, it is easy to show that the expressions (2), (3) and (4) are valid. These were normalised so that the same value of strain is present at the root, $\epsilon(0) = \epsilon_0$:

- first mode vibration of a uniform cantilevered beam ($k_1 \sim 1.875$):

$$w(\xi) = \frac{\epsilon_0}{2 k_1^2 \theta} \left[\cosh(k_1 \xi) - \cos(k_1 \xi) - \left(\frac{\cosh k_1 + \cos k_1}{\sinh k_1 + \sin k_1}\right)(\sinh(k_1 \xi) - \sin(k_1 \xi))\right] \qquad 2$$

- static deflection of a cantilever due to a point load at the tip, also approximating the dynamic deflection in the presence of a large point mass at the tip:

$$w(\xi) = \frac{\epsilon_0}{6\theta}(3\xi^2 - \xi^3) \qquad 3$$

- pure bending deflection of a uniform beam (boundary loads are such as to deform the beam into an arc):

$$w(\xi) = \rho - \sqrt{\rho^2 - \xi^2} \qquad \text{with: } \rho = \frac{\theta}{\epsilon_0} \qquad 4$$

The corresponding shapes are plotted in Figure 1 for θ=0.0125 and $\epsilon_0$ = 0.001, together with the tensile strain along the beam calculated, in each case, with expression (1). Note that in piezoelectric beams the tensile strain should be kept below 0.001 at all times to reduce the probability of early failure [26].

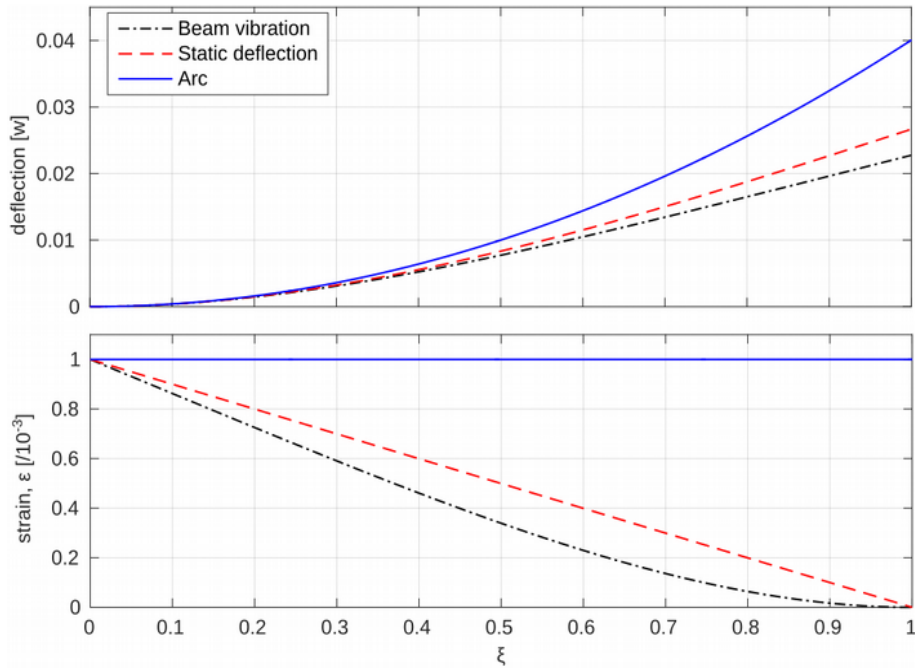

*Figure 1: Deflection of a beam in a selection of bending configurations and corresponding strains at the external convex surface.*

Observation of the figure reveals that whereas the strain in the arc is uniform, for a static deflection it decreases linearly from the maximum at the root to zero at the tip; for first mode of vibration, the strain decreases even faster. In other words, for shapes different from an arc the average strain is half (or less) of what it could be; the charges extracted from the material are correspondingly lower. Yet another point of view is to state that a design that uses the piezoelectric material more effectively by implementing pure bending may be able to provide the same harvested energy by using a third of the amount of material, which could mean reduced cost, mass and volume.

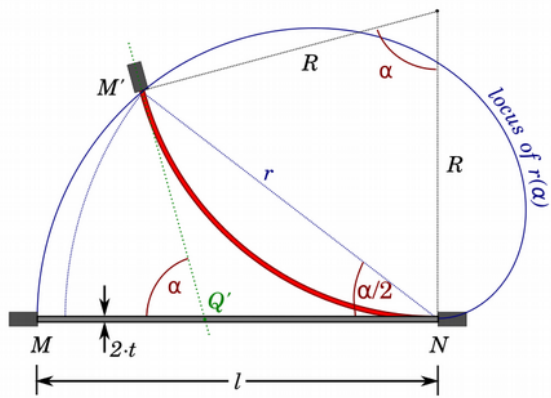

*Figure 2: Geometry of the beam bent into an arc*

Having established that deforming the beam into an arc is the most material-efficient route to piezoelectric energy harvesting, the objective is to devise an arrangement to produce the required end-loads. Figure 2 illustrates the relevant geometry, representing the beam bent while keeping the end at $N$ fixed (equivalently, we could think of M and N being rotated by equal and opposite angles). With reference to the figure for the meaning of the symbols, some key relations, derived from elementary geometry, are:

$$\epsilon = \frac{t}{R}$$

$$\alpha = \frac{l}{R}$$

$$r(\alpha) = \frac{2l}{\alpha} \sin\left(\frac{\alpha}{2}\right) \qquad 5$$

$$r(\epsilon) = \frac{2t}{\epsilon} \sin\left(\frac{l\epsilon}{2t}\right)$$

where ε is the maximum tensile strain present, found at the external surface of the beam. The most important observations are that the distance *r* between the tip (M) and the root (N) changes with deflection, implying that a simple rigid lever pivoted in N cannot be used to guide M along the correct path. Also, the tangent to the beam in M' intersects MN in a point Q' which moves along the beam with α, so neither that point is a good candidate as a pivot. Two key geometrical conditions must be satisfied: the ends rotate, assuming an angle α/2 from the line connecting them, and the distance between the ends reduces, according to expressions (5). The aim is to design a structure capable of satisfying, with good approximation, both requirements.

## Pure bending in a rotational device

In a rotational harvester like the Windmill [27] or the Pizzicato [22], piezoelectric bimorphs are cantilevered and placed radially, like spokes on a wheel; the relative rotation of external ring and central hub makes them vibrate or deflect. Neither device above attained both synchronicity and pure bending. The general problem stated in the introduction is therefore here modified for applications where the input energy takes the form of relative rotation. As discussed in the previous section, the structure must impose the same and opposite rotation on the two ends of the beam and accommodate the reducing distance between them. The structure must also permit the synchronous excitation of an array of bimorphs. The sketch of an arrangement with the potential of producing the required rotations is found in Figure 3. The bimorph is held between a rotating hub and an anchor which is hinged in P to an external ring, held fixed. A rigid rod connects a point A on the anchor to a point B on the hub, so that when the latter rotates, it forces a rotation of the anchor around P. Since this is only a sector of a circle, several bimorphs can be similarly mounted and simultaneously excited. With reference to the sketch for the labelling of the points, before rotation of the hub this equality is satisfied:

$$\vec{AB} = \vec{AP} + \vec{PO} + \vec{OB} \qquad 6$$

and after the hub has rotated by an angle φ:

$$\vec{A'B'} = \vec{A'P} + \vec{PO} + \vec{OB'} \qquad 7$$

The rotational requirement means: as the hub rotates clockwise (CW) by φ, the anchor holding the tip of the bimorph rotates counter-clockwise (CCW) by φ. This translates into the following transformations:

$$\vec{A'P} = \begin{bmatrix} \cos\phi & -\sin\phi \\ \sin\phi & \cos\phi \end{bmatrix} \vec{AP} \qquad 8$$

$$\vec{OB'} = \begin{bmatrix} \cos\phi & \sin\phi \\ -\sin\phi & \cos\phi \end{bmatrix} \vec{OB} \qquad 9$$

If the lengths of vectors $\vec{AB}$ and $\vec{A'B'}$ are equal, then an inextensible rod connecting points A and B will couple the two rotations to satisfy the requirement – when the hub rotates, the anchor will be forced to rotate by the same angle in the opposite direction.

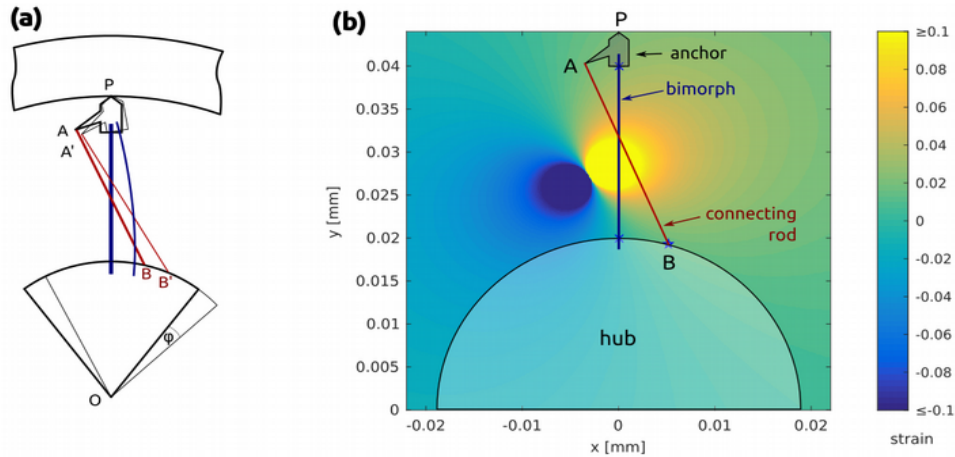

*Figure 3: (a) conceptual sketch of structure for rotational pure bending: thinner lines after rotation; (b) colormap of strain in the connecting rod according to location of end A.*

To explore the locus of the points where the above requirement is satisfied, a simple MATLAB script was developed which explores 'reasonable' locations for point A once both the required total rotation φ and the location of B are set. In principle, B could be anywhere, provided that it is fixed to the hub; it was selected as lying on its circumference to simplify manufacturing. It was observed that the locus changed with φ, particularly at large angles. Figure 3(b) illustrates a typical result, obtained with dimensions specific for the prototype developed later and for a rotation of 40 mrad (estimated to induce a uniform strain just over 0.1%)

In Figure 3(b), we notice two nearly circular regions where the strain rises steeply in absolute value to exceed 10%; care should be exercised to keep A far from these because of the extreme dependence of strain on exact location. Points in the upper-right region experience an elongation of the rod connecting A to B (positive strain), whereas a contraction is observed in the lower-left region (negative strain). For practical reasons, it is better to have A close to the anchor (as in Figure 3). In keeping with the current aim, the region near the anchor is within a uniform shade of colour just above 0% strain. These results apply to a situation which is in many ways ideal and not realisable in practice; nonetheless, they suggest that it is possible to find a location for point A such that the desired rotations of the beam's ends will leave $\|\vec{AB}\|$ roughly unchanged. When the direction of rotation is reversed, regions close to the anchor will be in compression (although there is no perfect symmetry). The resulting stress (positive or negative) will allow a connecting rod to develop the required forces. A final design should be symmetric, with criss-crossing rods.

As mentioned, the locus of points satisfying the condition $\|\vec{AB}\|=\|\vec{A'B'}\|$ changes with the angle of rotation; this is detrimental as it means that, as the beam deforms, it goes through stages when the strain is not uniform. However, it is observed that variations are not significant as long as rotations are limited to several degrees. This will be further addressed in the next section.

Results as in Figure 3 provide a starting point, but more accurate design needs to be performed with the aid of Finite Element Analysis (FEA). Corrections are needed to account for the compliance of the connecting rods, for example. Furthermore, more compliance needs to be built into the anchor, as necessary to accommodate the change in distance between the two ends of the beam during bending. Geometrical analyses also assume that all pivots are ideal, whereas the FE model is able to describe the behaviour of the compliant hinges that replace them in a real structure.

## FE modelling of rotational device

The structure designed to apply synchronous pure bending to a set of piezoelectric bimorphs is illustrated in Figure 4 (only one 45° sector of a full 8-element array is shown and modelled). The actual design is a 3D layered structure, where the bimorphs are sandwiched between two layers which provide the connecting rods (a prototype will be presented in the next section). The structure shown was designed for manufacture from sheet material. The plane strain approximation was used in the 2D FE model, because components that experience important strain are significantly thick, while they remain slender in the plane. The simple pivot between anchor and outside ring has been replaced by a compliant elliptical hinge, which permits rotation but also a certain degree of extension towards the hub. A technicality of the FE model is that the rods are modelled only in part and where they would cross and overlap the piezoelectric beam they are separately joined by connectors (mathematical constraints that behave like stiff connecting beams, by transferring forces and moments).

Figure 4(b) shows the longitudinal strain (along the vertical axis) in the bimorph as deformed by a rotation of 25 mrad. The colouration reveals that the strain is uniform along the beam and symmetric with respect to the neutral surface, as desired. The extremes of the colorbar (strain of ±0.13%) are observed only in the areas next to the clamps and are unlikely to appear in a real device where corners are less sharp and a thin layer of adhesive might be present. For not being central to the work, little effort has gone into optimising the reliability of the structure, nonetheless Figure 4(c) is reported to show that the von Mises stress in it (modelled as a 2 mm thick PMMA sheet) suggests a minimal risk of failure with less than 30 MPa observed everywhere but around a few FE nodes.

The uniformity along the beam of the longitudinal strain is even more manifest in the plots of Figure 5 (top row). The strain through the thickness of the beam is plotted for a selection of cross sectional lines along the beam. As the angle of rotation is increased from 12.5 to 100 mrad, the maximum strain observed increases as expected, still remaining fairly uniform along the beam (lines closely grouped) and symmetric with respect to the mid surface.

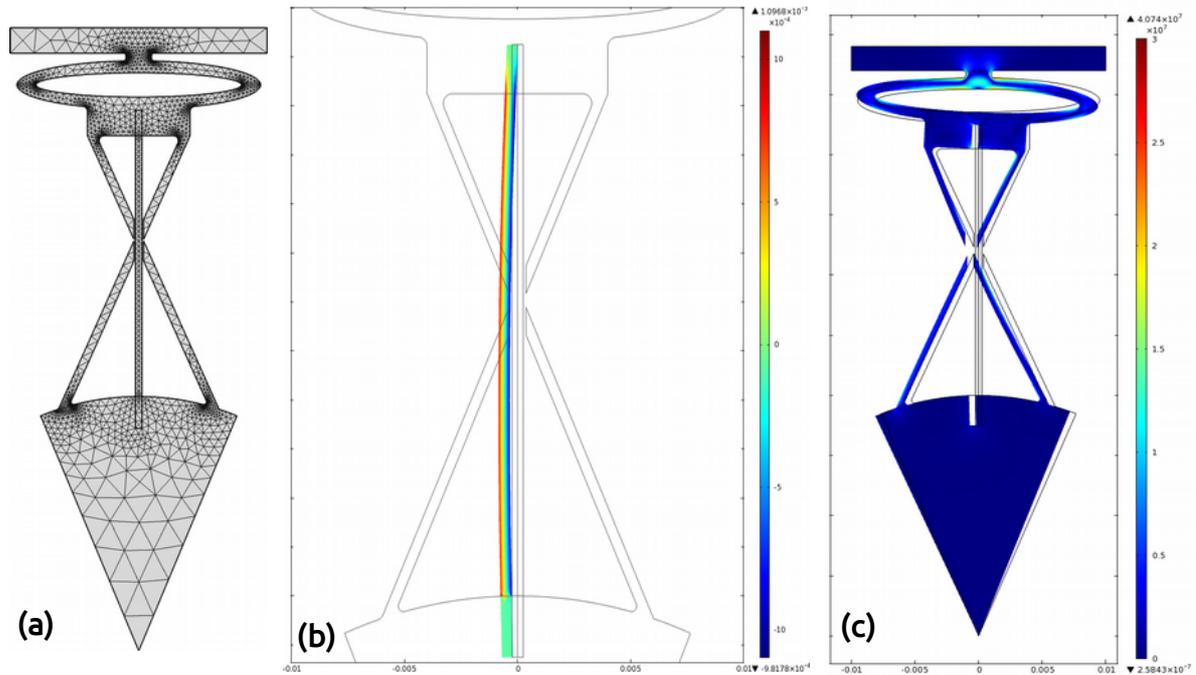

*Figure 4: (a) mesh of the FE model, from which the geometry can also be seen; solutions when α=25 mrad follow: (b) longitudinal strain (along the vertical axis) in the deformed beam, superimposed to the undeformed structure; (c) von Mises stress within the supporting structure.*

In summary, the results reported in this section demonstrate that it is possible to design a compliant smart structure than imparts pure bending deformation onto piezoelectric beams. A circular sector like this could be repeated several times (eight in the FE model above) and since the outer ring will be rigid, all beams will be deformed in synchronicity.

It is well known that a column in compression buckles into a shape which is not an arc but a half-sine; hence pure bending cannot be achieved by simply forcing the end of the beam to follow the locus *r(α)*. The strategy of the structure presented above is to prescribe the rotation of the ends, and leave the approach of the two ends as a simple consequence, by having a compliant anchor.

As a design guideline, it was observed that the stiffness of the connecting rod is key to a uniform strain: if the rods are too compliant, the anchor does not rotate sufficiently and lower strains are seen at that end; vice versa if lower strains are observed near the hub, the rods should be less stiff or point B (Figure 3) should be closer to the bimorph.

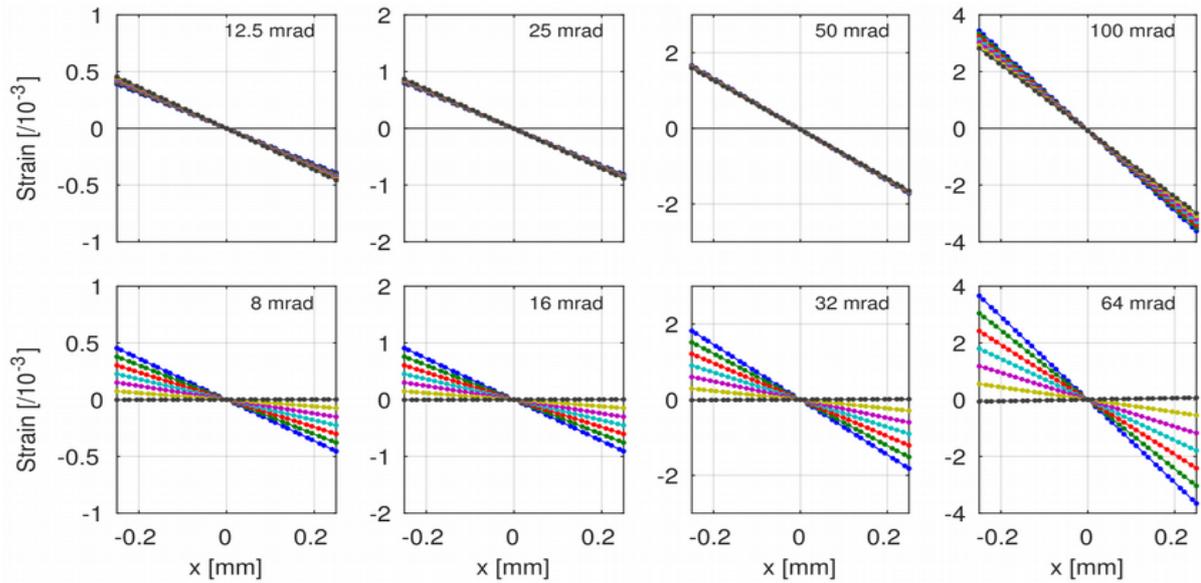

*Figure 5: plots of the longitudinal strain in the beam along parallel lines starting 1 mm above the hub and 3 mm apart; x is the transversal coordinate, with x=0 representing the middle of the beam. Top row: with connectors. Bottom row: without connectors. Angle of rotation indicated.*

The extreme situation is when the rods are completely removed, as in the lower row of plots in Figure 5. The strain progressively decreases moving from the hub to the external ring, confirming that the deflection is similar to that of a cantilever with tip load. Note that here the rotation in each case analysed has been set to 64% of the with-rods equivalent, to achieve comparable total charges. This design retains the advantage of synchronicity, but with approximately a third of energy harvesting capability (for comparable maximum strain). The connecting rods are an essential feature of the smart structure proposed.

*Table 1. Summary of total charge available and maximum strain at the outer surface in three positions along the beam bent into an arc (with connecting rods) and approximating a cantilever with tip loading (without rods), for a set of four rotation angles. The entries have been sorted by ascending charges, in view of energy harvesting applications.*

| Connecting rods | Angle [mrad] | Charge [µC] ↑ | Tensile strain near hub [$/10^{-4}$] | Tensile strain at middle [$/10^{-4}$] | Tensile strain near anchor [$/10^{-4}$] | Charge/angle ratio [µC/rad] |
|---|---|---|---|---|---|---|
| No | 8.0 | 0.72 | 4.54 | 2.26 | -0.02 | 90 |
| No | 16.0 | 1.44 | 9.09 | 4.52 | -0.04 | 90 |
| Yes | 12.5 | 1.46 | 3.89 | 4.22 | 4.55 | 117 |
| No | 32.0 | 2.88 | 18.2 | 9.04 | -0.13 | 90 |
| Yes | 25.0 | 2.90 | 7.97 | 8.34 | 8.71 | 116 |
| Yes | 50.0 | 5.69 | 16.6 | 16.26 | 15.9 | 114 |
| No | 64.0 | 5.71 | 36.6 | 18.00 | -0.67 | 89 |
| Yes | 100 | 10.99 | 34.4 | 31.34 | 28.2 | 110 |

Table 1 collects the surface strain calculated by FEA in three regions (near the hub, near the anchor and half way) and the corresponding total charges for a selection of conditions. Data confirm earlier statements that to generate the same amount of charge, the maximum strain in pure bending needs only be about half that at the root of a cantilever. The last column in the table, giving the ratio of charges over angle, is of interest for applications where the input angle is limited. Comparing the values with and without connectors, we notice that for the same rotation angle we obtain approximately 30% more charges with connectors than without.

## Prototype

A prototype was manufactured to verify the feasibility and the advantages of the ideas expressed thus far. As illustrated in Figure 6, it is composed of three layers: the external ones are laser cut from a 3 mm thick PMMA sheet and provide the connecting rods, whereas the middle layer holds the piezoelectric bimorphs. The smart structure is designed to accommodate up to 8 bimorphs, although, with no detriment to the objectives, only four were available to be installed. Unfortunately, two of these resulted permanently short circuited after assembly and therefore their signal could not be acquired. The bimorphs are parallel devices (Steminc, SMBA25W7T05PV) with an internal metal substrate of thickness 0.25 mm sandwiched between two piezo-active layers of thickness 0.125 mm. The material is a soft PZT branded SM411 (nominally equivalent to PZT-5J). The bimorphs are 7.1 mm wide and 25 mm long; approximately 2.5 mm of this length are embedded at either end for support, leaving 20 mm of active length. The two external electrodes were shorted together to form a two-terminal device with the substrate electrode. Two of these bimorphs, in diametrically opposite locations around the centre, were labelled PZT#2 and PZT#4 and became the focus of the measurements. PMMA was selected for anchors and connecting rods for its strain capabilities, availability and ease of manufacture via laser cutting. The faithfulness of the geometry to the FE model was limited by practical considerations and issues; in particular, details of anchors, rods and hinges are on the same scale as the accuracy of the laser cutting process.

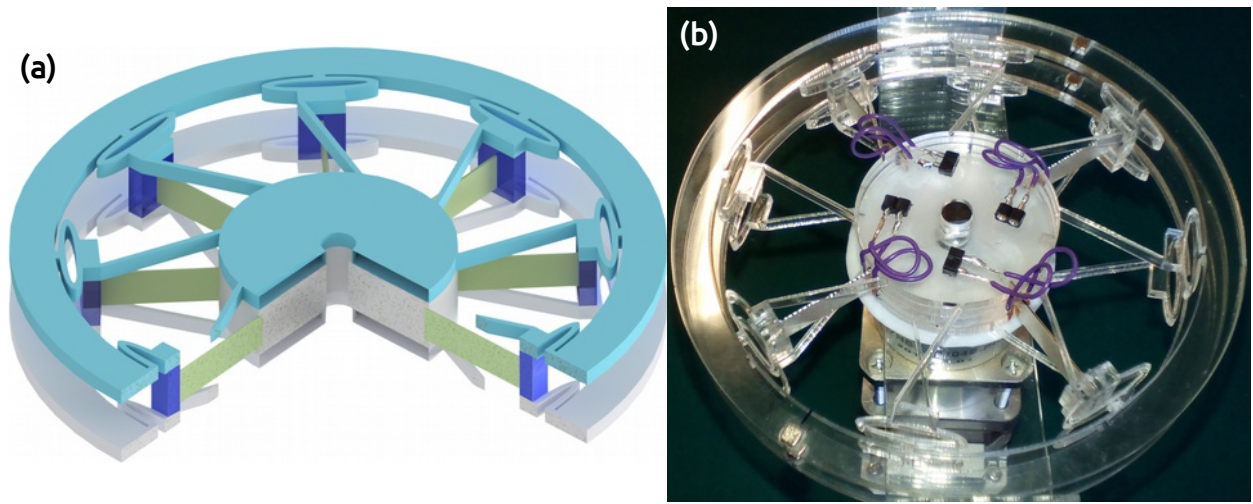

*Figure 6: (a) ¾ view of the harvester (b) photograph of the prototype.*

## Results

The first experiment measures the dependence of the charges produced by a bimorph on the angle of rotation. A torsional bar was fixed to the hub and moved between the limits of a series of four stoppers, designed to give controlled and reproducible maximum rotations. The angle of rotation in each direction was calculated from the linear displacement of the edge of the bar measured with a Dial Test Indicator, yielding an estimated uncertainty below 0.5 mrad. The charges flowing from one electrode to the other during each rotation were measured with an electrometer (Keithley 6517B); they are reproduced for one of the bimorphs in the top left graph of Figure 7. Note that the absolute values of the charges for the two directions of rotation are not equal due to the imperfect centring of bar and stopper. Figure 8 summarises the results for a selection of angles.

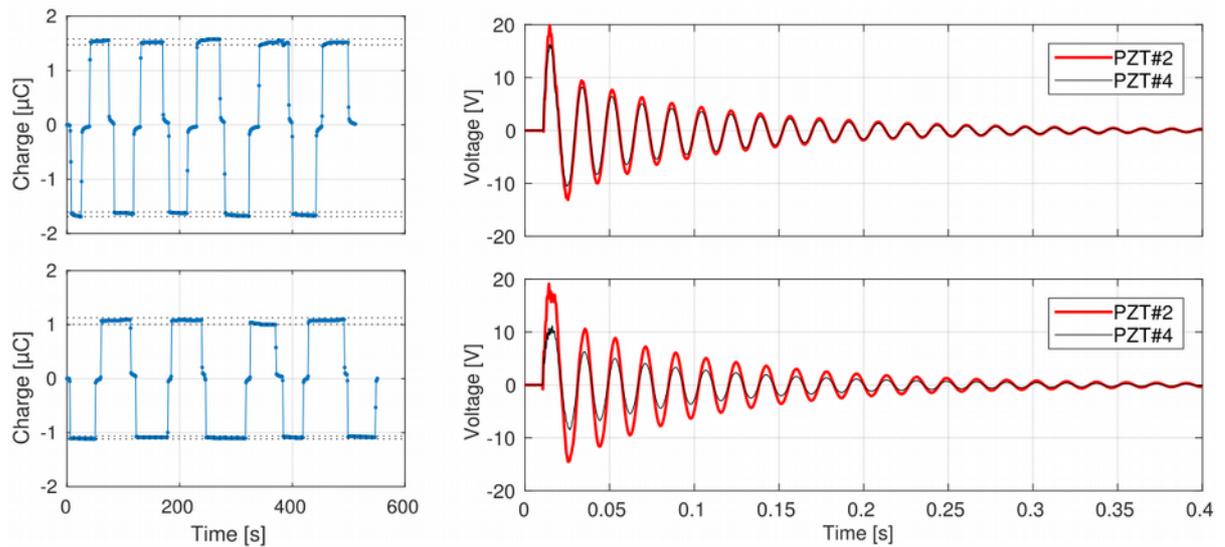

*Figure 7: (left column) charge vs. time in quasi-static tests. The hub was rotated alternatively CW and CCW, covering a total of 30.5 mrad, with brief pauses in the relaxed position; (right column) response of two bimorphs when prototype is subject to impact excitation; (top row) with connecting rods; (bottom row) after snapping-off connectors.*

The torsion bar was then struck to simulate impact excitation. Each bimorph was connected to a separate channel of the DAQ (NI 9221), so that the only impedance between its terminals was the input impedance of the DAQ (nominally 1MΩ//5pF). In all cases the two active bimorphs responded with similar amplitude and with the same frequency and phase. In Figure 7 one such event is plotted (top right), which demonstrates that synchronicity is achieved.

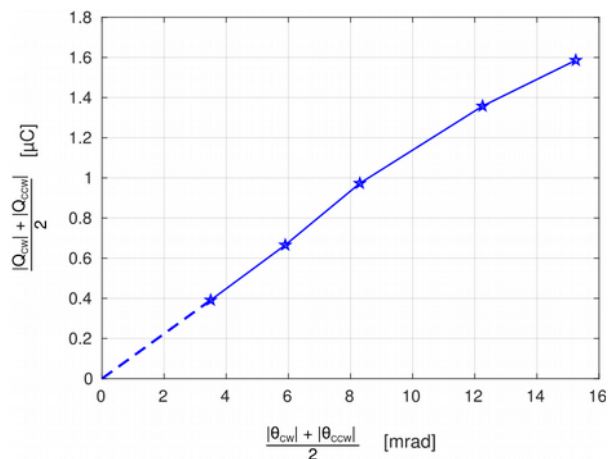

*Figure 8: charges vs. rotation angle for slow controlled rotation.*

## Analysis and Discussion

With the objective to determine if the bimorphs were bent into an arc or as tip-loaded cantilevers, macro photographs were taken of the beams under deflection and image processing applied to reconstruct their shape. Figure 9(a) reproduces a representative image: a Region of Interest was identified, converted from RGB to grayscale and then subtracted to its own copy shifted by one pixel in each direction to reveal the contour of the piezoelectric beam. The result was converted to black and white with suitable threshold to try to preserve only the contour of the beam (all this processing was performed manually within the GIMP software). The bitmap (.PNG) was then imported into MATLAB and the coordinate of each black pixel (defining the contour) were extracted into an array. The array was rotated in the 2D plane so that the interpolating line would be horizontal. This is for convenience, as only when placed horizontally (or vertically) it is possible to zoom in on the other direction and observe the curvature. At this point, the points defining the two edges of the beam were merged by translating one vertically by their average distance, to yield a single set of data. The resulting array *x* contained all the meaningful points available.

Fitting was performed with the `lsqnonlin` function of MATLAB 2015b, which finds the parameters *c* in a user-supplied function *f* which minimise the norm:

$$\min_{c} \|f(c; x)\| \qquad 10$$

In an implicit curve fitting application, *x* is the set of points and the function $f(c; x_i)$ gives the distance from a generic experimental point $x_i$ to the parametric curve. The optimisation is in the least-square sense.

As known, the equation that gives the static deflection of a cantilever with a load at the tip is

$$y = \frac{F}{EI}\left(\frac{Lx^2}{2} - \frac{x^3}{6}\right) \qquad 11$$

where *F* is the tip load, *L* the length, *E* the Young's modulus and *I* the second moment of area of the cross section. Since equation (11) makes clear distinction between the two coordinates and assumes the root in (0,0), it is valid only for a correctly aligned coordinate system and hence it is not suitable for fitting to the set of points extracted from a photograph. It was therefore generalised to allow the fitting to include roto-translations represented by:

$$\begin{bmatrix} x \\ y \end{bmatrix} = \begin{bmatrix} \cos\beta & -\sin\beta \\ \sin\beta & \cos\beta \end{bmatrix} \cdot \begin{bmatrix} x' - c_1 \\ y' - c_2 \end{bmatrix} \qquad 12$$

where $(c_1, c_2)$ are the coordinates of the root in the experimental data and β is the angle of rotation.

Therefore the function providing the vector whose norm is to be minimised is written (dropping the primes):

$$f(\vec{c}; x, y) = c_3 \left\{ \frac{L}{2}[(x-c_1)\cos c_4 - (y-c_2)\sin c_4]^2 - \frac{1}{6}[(x-c_1)\cos c_4 - (y-c_2)\sin c_4]^3 \right\} \\ -[(x-c_1)\sin c_4 + (y-c_2)\cos c_4] \qquad 13$$

where, additionally, $c_3$ = F/EI contains the curvature and $c_4$ is β, the (opposite of the) orientation of the tangent to the cantilever at the root. The location of the root was constrained to be within ±0.08 mm of the point that was identified on the photograph as representing the root. The range of ±0.08 mm is the based on the estimated uncertainty in the location of points.

Obviously, when fitting an arc of experimental points onto a circumference, rotation needs not be added, and the only fitting parameters are the centre $(a_1, a_2)$ and the radius $a_3$; the corresponding function is:

$$f(\vec{a}; x, y) = (x-a_1)^2 + (y-a_2)^2 - a_3^2 \qquad 14$$

The results of these analyses are summarised in Table 2 and in Figure 9(b). The table reports the fitting parameters as defined above and the mean norm, representing the average distance of each experimental point from the parametric curve. The mean norm was computed using the `dsearchn` function of MATLAB, applied between the experimental data and a very dense set of points computed from the fitting curve.

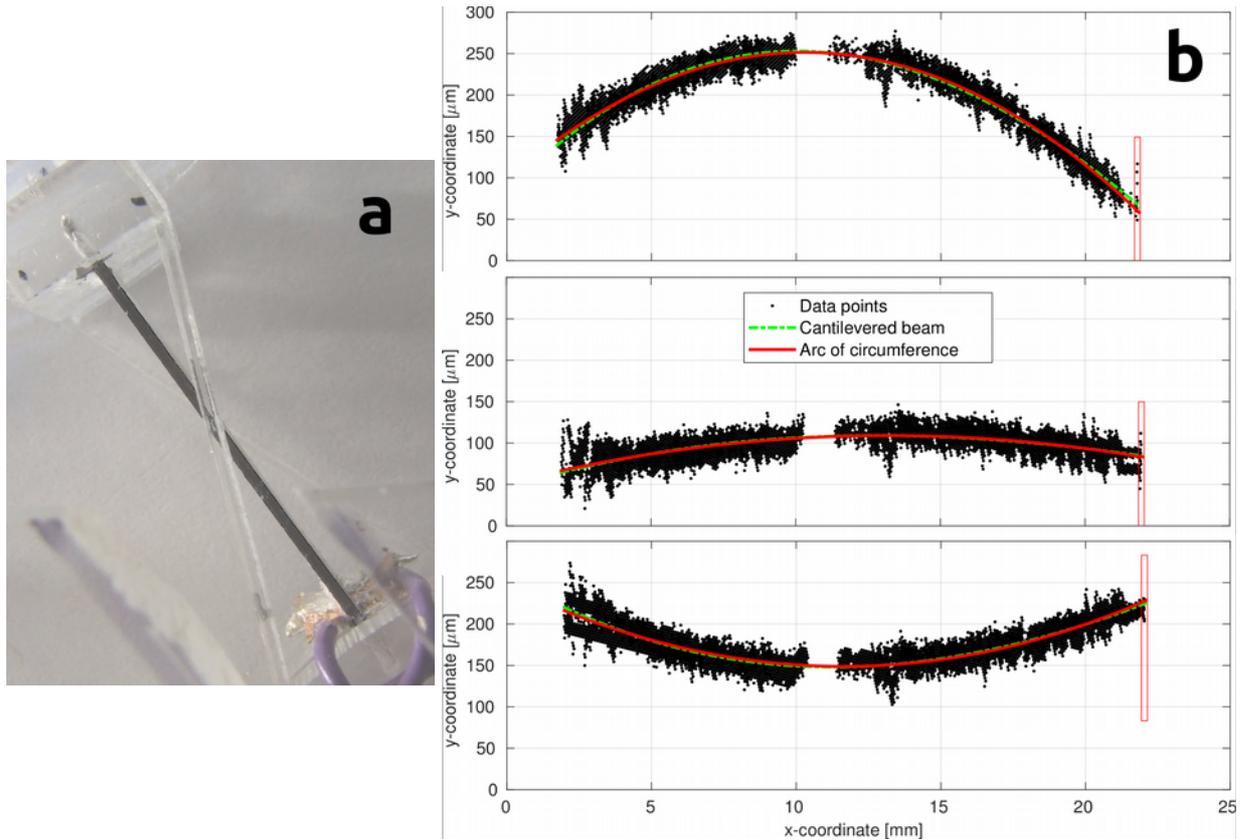

*Figure 9: (a) sample photograph with superimposed identified outline; (b) fitting of cantilevered beam deflection and arc to experimental points (dark pixels on the contour of (a)); top-down: convex, in rest position, concave deformation.*

Although the mean norms in the table are smaller for the arc than for the cantilever, they are too close to confidently conclude from them that the prototype structure is deflecting the bimorphs into arcs: the data points from the photographs are compatible with both shapes considered. In other words, the results just described are null, in the sense that they were unable to answer the research question that motivated them; they are reported for completeness.

*Table 2. Fitting parameters for the curves in Figure 9(b); refer to equations (13) and (14) for their meaning. The status is given with reference to the figure. The last column is the average distance of the experimental points to the fitted curve.*

| Status | $c_1$ [mm] | $c_2$ [mm] | $c_3$ [1/mm$^2$] | $c_4$ [mrad] | $a_1$ [mm] | $a_2$ [mm] | $a_3$ [mm] | norm [µm] |
|---|---|---|---|---|---|---|---|---|
| convex | 31.9 | 1.97 | -9.6E-5 | 29 | | | | 11.2 |
| | | | | | 20.3 | -344 | 346 | 10.8 |
| rest | 32.0 | 1.98 | -2.3E-5 | 5.1 | | | | 11.5 |
| | | | | | 23.2 | -1498 | 1500 | 11.2 |
| concave | 31.9 | 1.92 | 4.9E-5 | -13 | | | | 11.3 |
| | | | | | 21.6 | 692 | 690 | 11.3 |

It was therefore decided to snap off the connecting rods to measure the intrinsic gain produced by their presence. Data plotted in Figure 7 and partly already discussed, were collected just before and just after such operation, to ensure experimental conditions were not altered. Each of the graphs on the right reproduces the signal from two bimorphs under impact excitation, as described earlier. Using the same labels as in the legend, PZT#2 was preserved intact, whereas PZT#4 had the connecting rods snapped off. Whereas synchronicity is preserved, as expected, it is clear that the signal from PZT#4, already lower for some intrinsic difference between bimorphs or mounting, is significantly reduced by the removal of the connectors. As the interest is in energy harvesting, it is useful to look at the ratio of energies produced by the two devices:

$$R_{energy} = \frac{\int V^2_{PZT\#4}(t)}{\int V^2_{PZT\#2}(t)} \quad\quad 15$$

For the impact events in the figure, such ratio is 0.73 with connectors and 0.33 after their removal. This means that the PZT#4 has lost about 55% of its energy generation capability. The two graphs on the left represent the variation of charge flowing through the external circuit of PZT#4 as a function of time when the torsional bar is alternatively moved between the extremes of travel defined by the stopper. In both cases, the overall rotation of the torsional bar covers 30.5 mrad. Statistical analysis of these data yields a total charge difference of 3.2±0.1 µC with rods and 2.2±0.1 µC once rods were snapped off, implying a loss of 31% or, alternatively, that the connectors boosted performance by 45%.

## Conclusions

A novel compliant smart structure for kinetic piezoelectric energy harvesting has been presented, geometrically analysed, modelled with FE and prototyped. The structure, featuring a set of bimorphs, achieves the objectives of providing uniform strain along each bimorph and synchronicity of deflection within the set. In this way, both suboptimal utilisation of piezoelectric material and signal cancellation due to out-of-phase generation, common in previous harvesters, are successfully eliminated.

The FE model proves that the proposed structure delivers uniform strain within the bimorphs and that the connecting rods are essential to this success (Figure 5).

The prototype demonstrates synchronicity of the generated power signals. The analysis of photographs yielded a null result, being compatible with both hypotheses considered (arc and cantilevered beam). However, the removal of the connecting rods has demonstrated that, in line with FEA, they offered a significant boost in performance, supporting the conclusion that pure bending was indeed achieved in the prototype (Figure 7).

Data in the last column of Table 1 show that without connecting rods a rotation ~30% larger is needed to generate the same charges. Already, this means that their presence reduces the input displacement requirements of the harvester. More importantly, though, the peak strain in the material is twice larger without connectors, with serious negative impact on service life. Alternatively, we can conclude from Table 1 that for a set maximum strain, and hence design life, we can obtain over twice as many charges with connecting rods, and the pure bending deformation they afford. As energy harvesters move towards applications, their reliability over time must receive more attention. The structure presented offers the best use of the material available, within its operating limits.

This smart structure accepts reciprocating rotations of a few degrees as mechanical input. These may be originally oscillatory rotations or they could be obtained by transforming linear vibrations into rotational ones. Alternatively, energy can be input into the structure via impact (as in Figure 7) or via plucking.

# References

[1]   Samson D, Kluge M, Becker T and Schmid U 2011 Wireless sensor node powered by aircraft specific thermoelectric energy harvesting *Sens. Actuators Phys.* **172** 240–4

[2]   Naruse Y, Matsubara N, Mabuchi K, Izumi M and Suzuki S 2009 Electrostatic micro power generation from low-frequency vibration such as human motion *J. Micromechanics Microengineering* **19** 094002

[3]   Ylli K, Hoffmann D, Willmann A, Becker P, Folkmer B and Manoli Y 2015 Energy harvesting from human motion: exploiting swing and shock excitations *Smart Mater. Struct.* **24** 025029

[4]   Priya S 2007 Advances in energy harvesting using low profile piezoelectric transducers *J. Electroceramics* **19** 167–84

[5]   Ferrari M, Ferrari V, Guizzetti M, Marioli D and Taroni A 2008 Piezoelectric multifrequency energy converter for power harvesting in autonomous microsystems *Sens. Actuators Phys.* **142** 329–35

[6]   Marin A, Turner J, Ha D S and Priya S 2013 Broadband electromagnetic vibration energy harvesting system for powering wireless sensor nodes *Smart Mater. Struct.* **22** 075008

[7]   Challa V R, Prasad M G, Shi Y and Fisher F T 2008 A vibration energy harvesting device with bidirectional resonance frequency tunability *Smart Mater. Struct.* **17** 015035

[8]   Ferrari M, Baù M, Guizzetti M and Ferrari V 2011 A single-magnet nonlinear piezoelectric converter for enhanced energy harvesting from random vibrations *Sens. Actuators Phys.* **172** 287–92

[9]   Sato T and Igarashi H 2015 A chaotic vibration energy harvester using magnetic material *Smart Mater. Struct.* **24** 025033

[10]   Wood R J, Steltz E and Fearing R S 2005 Optimal energy density piezoelectric bending actuators *Sens. Actuators Phys.* **119** 476–88

[11]   Goldschmidtboeing F and Woias P 2008 Characterization of different beam shapes for piezoelectric energy harvesting *J. Micromechanics Microengineering* **18** 104013

[12]   Guan Q C, Ju B, Xu J W, Liu Y B and Feng Z H 2013 Improved strain distribution of cantilever piezoelectric energy harvesting devices using H-shaped proof masses *J. Intell. Mater. Syst. Struct.* **24** 1059–66

[13]   Xu J W, Liu Y B, Shao W W and Feng Z 2012 Optimization of a right-angle piezoelectric cantilever using auxiliary beams with different stiffness levels for vibration energy harvesting *Smart Mater. Struct.* **21** 065017

[14]   Paquin S and St-Amant Y 2010 Improving the performance of a piezoelectric energy harvester using a variable thickness beam *Smart Mater. Struct.* **19** 105020

[15]   Zheng Q and Xu Y 2008 Asymmetric air-spaced cantilevers for vibration energy harvesting *Smart Mater. Struct.* **17** 055009

[16]  Becker P, Hymon E, Folkmer B and Manoli Y 2013 High efficiency piezoelectric energy harvester with synchronized switching interface circuit *Sens. Actuators Phys.* **202** 155–61

[17]  Goldacker W, Schlachter S I, Nast R, Reiner H, Zimmer S, Kiesel H and Nyilas A 2002 Bending strain investigations on BSCCO(2223) tapes at 77 K applying a new bending technique *AIP Conf. Proc.* **614** 469–76

[18]  Takayasu M, Chiesa L, Harris D L, Allegritti A and Minervini J V 2011 Pure bending strains of Nb 3 Sn wires *Supercond. Sci. Technol.* **24** 045012

[19]  Eguizabal J, Tufaga M, Scheer J K, Ames C, Lotz J C and Buckley J M 2010 Pure moment testing for spinal biomechanics applications: Fixed versus sliding ring cable-driven test designs *J. Biomech.* **43** 1422–5

[20]  Lysack J T, Dickey J P, Dumas G A and Yen D 2000 A continuous pure moment loading apparatus for biomechanical testing of multi-segment spine specimens *J. Biomech.* **33** 765–70

[21]  Boers S H A, Geers M G D and Kouznetsova V G 2010 Contactless and Frictionless Pure Bending *Exp. Mech.* **50** 683–93

[22]  Pozzi M 2016 Magnetic plucking of piezoelectric bimorphs for a wearable energy harvester *Smart Mater. Struct.* **25** 045008

[23]  Pozzi M, Almond H J, Leighton G J T and Moriarty R J 2015 Low-profile and wearable energy harvester based on plucked piezoelectric cantilevers *Proc. SPIE Smart Sensors, Actuators, and MEMS VII; and Cyber Physical Systems* vol 9517 (Barcelona (Spain)) pp 951706-951706–9

[24]  Pozzi M and Zhu M 2011 Plucked piezoelectric bimorphs for knee-joint energy harvesting: modelling and experimental validation *Smart Mater. Struct.* **20** 055007

[25]  Elliott A D and Mitcheson P D 2012 Implementation of a single supply pre-biasing circuit for piezoelectric energy harvesters *Procedia Eng.* **47** 1311–1314

[26]  Guillon O, Thiebaud F and Perreux D 2002 Tensile fracture of soft and hard PZT *Int. J. Fract.* **117** 235–46

[27]  Priya S 2005 Modeling of electric energy harvesting using piezoelectric windmill *Appl. Phys. Lett.* **87** 184101